\documentclass[aps,amsmath,amssymb,nofootinbib]{revtex4}

\usepackage[pdftex]{graphicx}
\usepackage{graphics}
\usepackage{amsmath}
\usepackage{subfigure}

\usepackage{verbatim}
\usepackage{color}
\usepackage{multirow}
\usepackage{hyperref}
\usepackage{enumitem}
\usepackage{float}

\newcommand{\dst}{\displaystyle}

\begin{document}
\title{A Soliton and a Black Hole are in Gauss-Bonnet gravity. Who wins?}

\author{Anson W.C. Wong$^{1,2}$}
\email{awcwong@physics.ubc.ca}
\author{Robert B. Mann$^{1,3}$}
\email{rbmann@perimeterinstitute.ca}
\address{$^{1}$ Department of Physics and Astronomy, University of Waterloo, Waterloo, Ontario, N2L 3G1, Canada}
\address{$^{2}$ Department of Physics and Astronomy, University of British Columbia, 6224 Agricultural Road, Vancouver, B.C., V6T 1W9, Canada}
\address{$^{3}$ Perimeter Institute for Theoretical Physics, Waterloo, Ontario, N2L 2Y5, Canada}

\begin{abstract}
We study here the phase-transitional evolution between the Eguchi-Hanson soliton, the orbifolded Schwarzschild Anti de-Sitter black hole, and orbifolded thermal Anti de-Sitter space in Gauss-Bonnet gravity for a small Gauss-Bonnet coefficient $\alpha$.
Novel phase structure is uncovered for both negative and positive $\alpha$ with spacetime configurations that are stable in Gauss-Bonnet gravity without being so in Einsteinian gravity.
The evolutionary tracks taken towards such stable configurations are guided by quantum tunnelling and can be represented with a phase diagram constructed by comparing the Euclidean actions of each of our states as a function of $\alpha$ and the black hole radius $r_b$.
According to the AdS/CFT correspondence dictionary, it is expected that some generalized version of closed-string tachyon condensation will exhibit the phase behaviour found here.
\end{abstract}

\maketitle

\section{Introduction}

One of the more celebrated milestones of theoretical physics is the Anti-de Sitter/Conformal Field Theory (AdS/CFT) correspondence conjectured by Maldecena \cite{Maldacena1999-AdSCFT}.
In a nutshell, the conjecture posits that the conformal boundary of an $n$-dimensional gravitational theory in Anti de-Sitter (AdS) space has a mathematically equivalent $(n-1)$-dimensional quantum field theory.
By calculating realizable quantities on the gravitational side of this correspondence, much insight can be grasped from its field theory dual without making explicit calculations within the framework of the field theory itself and vice versa.

An early example of this is the Hawking-Page phase transition between thermal AdS and the Schwarzschild AdS black hole \cite{Hawking1982-HP_transition} and the corresponding confinement/deconfinement phase transition in the large $N$ limit of ${\cal N}=4$ super Yang-Mills theory \cite{Witten1998-AdS_Holography_confinement_defconfinement}.  More recent examples are associated with a conjecture that closed string tachyon condensation
is dual to a soliton configuration (or `bubble of nothing') \cite{Horowitz2006-Quasilocal_Tachyons}, such as  the AdS solitons obtained by Horowitz and Myers \cite{Horowitz1999-AdS_soliton}.  An interesting avenue of investigation along these lines is the study of three asymptotically-AdS$/\mathbb{Z}_k$ spacetimes together: the Eguchi-Hanson soliton, the orbifolded Schwarzschild AdS black hole, and orbifolded thermal AdS space. These objects exist on the gravitational side of the correspondence and can tell an entertaining story in the context of their phase transitions amongst each other; but before we explain why this is so, it is convenient to first briefly introduce our three said spacetimes.

\emph{Anti de-Sitter spacetime} is often viewed as a spacetime corresponding to a negative cosmological constant in the standard Friedmann cosmology that string theories naturally compactify on \cite{Prokopec2011-negative_cosmo_constant_AdS}.
Physically-interesting candidate supergravity models for quantum gravity typically admit solutions that are asymptotic to this spacetime.
Modding out by $\mathbb{Z}_k$ yields the orbifolded anti de-Sitter spacetime  (OAdS).
In relation to this work, OAdS is very much treated as a `background' spacetime, which is to say that both the AdS black hole and the Eguchi-Hanson soliton converge towards it in the large distance $r \rightarrow \infty$ limit.

The (orbifolded) \emph{Schwarzschild-AdS black hole} (OSAdS) is a statically uncharged black hole embedded in (orbifolded) AdS space.
It is spherically symmetric and has an event horizon at some $r=r_b$ where events within $r < r_b$ cannot affect external observers situated outside of the event horizon at $r > r_b$.
For an AdS cosmological length of $\ell$, the mass of such a black hole takes the simple expression $M_b = r_b^{n-3}\left( 1+r_b^2/\ell^2 \right)$.
Accompanying this is the no-hair theorem which reveals how this mass fully characterizes the OSAdS because of its constant (zero) angular momentum and charge, thereupon leaving its mass as the only degree of freedom left.
Semi-classical reasoning would suggest that the OSAdS dissipates into the aforesaid OAdS by the means of Hawking radiation.

Lastly, the \emph{Eguchi-Hanson soliton} (EHS) obtained by Clarkson and Mann \cite{Clarkson2006-OddSoliton}  (and independently Copsey \cite{Copsey2007-Bubbles2}) is perturbatively of lowest energy within its asymptotically-AdS$/\mathbb{Z}_k$ class and thereby considered a ground state within string theory.
Closed-string tachyon condensation is the expected field theory dual process corresponding to phase transitions from EHS  to the OSAdS and OAdS geometries.
The EHS is an example of an expanding `bubble of nothing' that serves to decay the stabilizing potential over time by reducing redundant dimensions and leaving behind an AdS spacetime with a curvature approaching zero \cite{Horowitz2006-Quasilocal_Tachyons,Horowitz2005-Tachyon_Condensation,Hikida2007-dual_AdSorbifolds,Copsey2007-Bubbles2,Brown2012-bubble_of_nothing}.

All three of the aforementioned spacetimes have non-trivial topology insofar as 
their ambient AdS spacetimes  are quotiented by the isometric group $\mathbb{Z}_k$ (integers modulo $k$). For the EHS, this procedure  smooths out what would otherwise be a conically-singular spacetime.  For the other two spacetimes this procedure is called orbifolding;
it can be thought of as a folding-up of the manifold to reduce its large fundamental domain. 
The field theory counterpart to orbifolding is the formation of a fixed point at the cone-like center of the AdS orbifold such that closed-string tachyons (field theoretic objects dual to general relativistic singularities) locally wind around it to only allow for normalizable modes \cite{Silverstein2006_Singularities_ClosedStringTachyons,Hikida2007-dual_AdSorbifolds}.

We are now ready to discuss the concept of phase transitions amongst the EHS, OSAdS, and OAdS states.
One way to portray this is by drawing a phase diagram comprised of individual phase regions; each encoding the directions of quantum tunnelling between every pairwise combination of the EHS, OSAdS, and OAdS spacetimes.
To that end, a phase region prescribes an evolutionary track for each of our states to take to reach a state of stability based on relative energetics.
By adjusting the black hole temperature $T_b$ (corresponding to varying the black hole radii $r_b$), one can traverse through the phase diagram and reach different phase regions by crossing the phase boundaries that divide them.

It is because of the interesting phase transitions that occur between the EHS, OSAdS, and OAdS states which makes for an equally interesting field theory dual.
For a case in point, the previously mentioned Hawking-Page phase transition between the OSAdS with OAdS at critical temperature corresponds to the quark confinement/deconfinement phase transition in the large $N$ limit of $\mathcal{N}=4$ super Yang-Mills theory \cite{Hawking1982-HP_transition}.
This was first proposed by Witten \cite{Witten1998-AdS_Holography_confinement_defconfinement} from an inspection that the low temperature confinement phase corresponded to AdS space, and the high-temperature deconfinement phase corresponded to the AdS black hole.
The gravity-side phase transitions between the EHS, OSAdS, and OAdS have been studied in Einsteinian gravity for the odd dimensions relevant to string theory \cite{Stotyn2009-soliton_bh_phasetransitions} with a belief that its corresponding field theory dual is the condensation of localized closed-string tachyons \cite{Hikida2007-dual_AdSorbifolds,Hikida2006-phasetransitions_largeN_gauge}.
What interests us is the question of how all this extends to a theory of gravity that is more general than Einsteinian gravity.

Although conceptually important, Einsteinian gravity fails to recognize the free-field dynamics in the low-energy sector for $n \ge 5$ dimensional superstring theories as it ignores higher-order quantum loop corrections \cite{Dadhich2008-Characterization_of_Lovelock}.
The inherent self-interactive nature of gravity brings in higher-order curvature terms into the field equations that are physically realizable only when the dimensionality is high enough.
A promising remedy for this problem is Lovelock gravity \cite{Lovelock1971-LG1,Lovelock1970-LG2,Lovelock1975-LG3}.
This gravitational theory adds a unique linear combination of higher-order Riemann curvature correction terms to the Einsteinian action {\it without} introducing the quantum instabilities, also known as ghosts, prevalent in many higher-order gravitational theories.
In $n=5$ dimensions, Lovelock gravity is called Gauss-Bonnet (GB) gravity and is the simplest case of Lovelock gravity where the higher-order curvature corrections can be controlled fully with a single parameter $\alpha$. 
The GB version of the OSAdS has been well-studied throughout the last decade \cite{Cai2002-GB_BH_AdS} but the GB version of the EHS has only been found recently via semi-analytic and numerical methods by Wong and Mann \cite{Wong2012-GB_soliton}, hence making it natural (and now possible) to investigate how phase transitions between EHS, OSAdS, and OAdS differ upon considerations of the stringy corrections induced by GB gravity.
Finding new phase transitions will infer finding extra phase behavior from its field theory dual as a perk.

The GB phase transitional work between the OSAdS and thermal OAdS is not a new concept \cite{Cai2002-GB_BH_AdS,Cho2002-AdS_GB_phase,Anninos2009-Thermodynamics_GB_AdS_BH_higherderivatives}, but our additional consideration of the EHS into the mix is.
We expect promising results to come of this $\alpha \neq 0$ study given that we already have EHS being the ground state in $\alpha=0$ gravity for small black hole radii (and not OAdS if we naively studied only the OSAdS and OAdS states).
The goal of this paper is to {\it qualitatively} study the GB phase transitions between the $5$-dimensional EHS, OSAdS, and OAdS as a function of a small Gauss-Bonnet parameter $\alpha$ and $r_b$.

The outline of the paper will be as follows.
In Section \ref{sec:gaussbonnet_gravity}, we introduce the reader to GB gravity.
Afterwards, the definitions and details of the EHS, OSAdS, and OAdS states in GB gravity will be provided in Section \ref{sec:EHS_OSAdS_OAdS}.
By comparing the actions of these states, the corresponding evolutionary tracks on each phase region in the phase diagram can be found; such details will be explained in Section \ref{sec:phase_transitions_and_evolutionary_tracks}.
How the set of evolutionary tracks varies with $\alpha$ is analyzed in Section \ref{sec:resuts_and_discussion} with discussions and comparisons of its novel phase transitions, or the lack thereof, with that of Einsteinian gravity.
Lastly, we summarize our results and note  prospects for future work in Section \ref{sec:summary_and_prospective}.

\section{Gauss-Bonnet Gravity: Lovelock gravity in $n=5$ dimensions}
\label{sec:gaussbonnet_gravity}
We will consider only $n=5$ dimensional spacetimes for the rest of this paper.
In $n=5$ dimensions, Lovelock gravity differs from Einsteinian gravity by no more than a set of Riemann curvature squared terms in the Lagrangian.
Mathematically, this is because the action is a summation of dimensionally extended Euler densities, and in $n=5$ dimensions, the higher-than-cubic curvature terms become topologically trivial.
Although our study concerns itself with only Gauss-Bonnet gravity, we will introduce Lovelock gravity in full generality to begin with.
The Lovelock Lagrangian $\mathcal{L}$ can be written as a sum of Euler densities 
\begin{equation}
\mathcal{L} =  \sqrt{-g} \sum^{\infty}_{k=0} c_k \mathcal{L}_k = \sqrt{-g} \left[ c_0 \mathcal{L}_0 + c_1 \mathcal{L}_1 + c_2 \mathcal{L}_2 + c_3 \mathcal{L}_3 + ... \right] ,
\label{eqn:lovelock_action}
\end{equation}
where $\sqrt{-g}$ is the manifold volume element calculated using the determinant $g$ of the metric.
Each Euler density $\mathcal{L}_k$ is defined by a unique linear combination of Riemann curvature terms of exactly order $k$
\begin{equation}
\mathcal{L}_k \equiv \frac{1}{2^k} \bar{\delta}^{a_1 ... a_k b_1 ... b_k}_{c_1 ... c_k d_1 ... d_k} R^{c_1 d_1}_{a_1 b_1} ... R^{c_k d_k}_{a_k b_k} , 
\label{eqn:lovelock_k_action}
\end{equation}
where the rank-$(4k)$ tensor $\bar{\delta}$ is defined as the normalized antisymmetric product of Kronecker delta functions
\begin{equation}
\bar{\delta}^{a_1 ... a_k b_1 ... b_k}_{c_1 ... c_k d_1 ... d_k} \equiv \frac{1}{n!} \delta^{a_1}_{[ c_1} \delta^{b_1}_{d_1} ... \delta^{a_k}_{c_k} \delta^{b_k}_{d_k ]} .
\label{eqn:delta_tensor}
\end{equation}

The total action $\mathcal{I}$ is
\begin{equation}
\mathcal{I} = - \frac{1}{16 \pi G} \int d^{n} x \sqrt{-g} \left[ c_{0} \mathcal{L}_{0} + c_{1} \mathcal{L}_{1} + c_{2} \mathcal{L}_{2} + c_{3} \mathcal{L}_{3}  + ... \right] ,
\label{eqn:GB_action_raw}
\end{equation}
 where $G$ is the $n$-dimensional gravitational constant.

As discussed earlier, the topological triviality of the Euler densities implies that only the first $\mathcal{L}_0,\mathcal{L}_1,...,\mathcal{L}_{\left[ n/2 \right]}$ terms in \eqref{eqn:GB_action_raw} are relevant in $(n+1)$ dimensions
(where $\left[ \cdot \right]$ is the floor function), meaning that the knowledge of $\mathcal{L}_0$, $\mathcal{L}_1$, and $\mathcal{L}_{2}$ will suffice for GB gravity.
The evaluation of these give
\begin{equation}
\begin{array}{ccccc}
\mathcal{L}_0 = 1, & {} & \mathcal{L}_1 = R, & {} & \mathcal{L}_{\text{GB}} \equiv \mathcal{L}_2 = R_{abcd}R^{abcd} - 4R_{ab}R^{ab} + R^2 .
\end{array}
\label{eqn:GB_action_terms}
\end{equation}

By setting the zeroth and first order coefficients of the action \eqref{eqn:GB_action_raw} as $c_{0} = -2 \Lambda$ and $c_{1} = 1$ to match Einsteinian gravity at $\alpha=0$, the (Euclidean) Einstein-Gauss-Bonnet (bulk) action becomes
\begin{equation}
\mathcal{I}_{\text{GB}} = - \frac{1}{16 \pi \text{G}} \int d^{n} x \sqrt{-g} \left[ -2 \Lambda + R + \alpha \mathcal{L}_{\text{GB}} \right] ;
\label{eqn:GB_action_final}
\end{equation}
an integration over $1$ timelike component and $(n-1)$ spacelike components of our metric.
We have introduced the notation $\alpha=c_{2}$ as the sole Gauss-Bonnet coefficient used for tuning the relative magnitude of the curvature-squared (quantum) correction terms.

In general, there are $r=\infty$ boundary terms attached to the action \eqref{eqn:GB_action_final} but we can excuse ourselves for not including them here because the metrics of interest here converge to the same thermal AdS background at infinity, thus leaving physically-relevant quantities (such as the relative action) to depend only on the difference between   bulk quantities.
Although $\mathcal{I}_{\text{GB}}$ is generally divergent for asymptotically AdS spacetimes because of its infinite spatial volume, the relative actions can remain finite if the metrics share the same asymptotics and have certain coordinate periodicities matched-up properly -- a more precise description of this procedure will be discussed in the main text.

Upon extremizing the GB action \eqref{eqn:GB_action_final} with respect to the metric, we obtain the GB vacuum field equations
\begin{equation}
\begin{array}{lcll}
\Bigl[ R_{m n} - \frac{1}{2}Rg_{m n} + \Lambda g_{m n} \Bigr] & - &
\alpha \Bigl[ \ \frac{1}{2}g_{m n}(R_{abcd}R^{abcd} - 4R_{ab}R^{ab} + R^2) - 2RR_{m n} & \\ & & \ \ \ \ \ + 4R_{m a}R^{a}_{n} + 4R^{a b}R_{m a n b} - 2R^{abc}_{m}R_{n a b c} \ \ \ \ \ \ \ \ \ \Bigr] = 0 .
\end{array}
\label{eqn:GB_fieldequations}
\end{equation}
The GB field equations \eqref{eqn:GB_fieldequations} show explicitly the contribution of the Gauss-Bonnet terms toward the pure Einstenian vacuum field equations.

\section{The Eguchi-Hanson Soliton, Black Hole, and thermal AdS in Gauss-Bonnet gravity}
\label{sec:EHS_OSAdS_OAdS}

We will now introduce the GB EHS, GB OSAdS, and OAdS spacetimes with the occasional use of state notation $i \in \mathcal{S} = \{ s \equiv \text{EHS}, b \equiv \text{OSAdS}, a \equiv \text{OAdS} \}$.
All three metrics admit dimensionless solutions $f_i(r,\alpha)$, $g_i(r,\alpha)$, $h_i(r,\alpha)$ that solve the GB field equations \eqref{eqn:GB_fieldequations} with the ansatz
\begin{equation} 
ds^2 = 
-\frac{r^2}{\ell^2} g_i(r,\alpha) dt^2 + 
\frac{\ell^2}{r^2} \frac{dr^2}{h_i(r,\alpha) f_i(r,\alpha)} + 
\frac{r^2}{4} \left( d\theta^2 + \sin^2 \theta d\phi^2 \right) + 
\frac{r^2}{4} f_i(r,\alpha) \left(d\psi_i + \cos \theta d\phi \right)^2 ,
\label{eqn:metrics} 
\end{equation} 
where $\ell$ is the AdS cosmological length.

The range of the timelike, radial, and three angular coordinates are $t \in \left(-\infty,\infty \right)$, $r \in \left[ r_i,\infty \right)$, $\theta \in \left[ 0,\pi \right]$, $\phi \in \left[ 0,2\pi \right]$, and $\psi_i \in \left[ 0,\eta_i \right]$, respectively.
Once the orbifold-allowed values for $\eta_i$ are set, these types of spacetimes are asymptotic 
to OAdS provided the condition that $f_i$, $g_i$, and $h_i$ are  convergent to unity in the $r \rightarrow \infty$ limit is imposed.
In the following subsections, the EHS, OSAdS, and OAdS spacetimes in GB gravity will be introduced in terms of their solution functions $f_i$, $g_i$, and $h_i$.

\subsection{Gauss-Bonnet Eguchi-Hanson Soliton}
\label{subsec:GB_EHS}

The 5-dimensional Einsteinian $\alpha=0$ EHS has the metric 
\begin{equation}
ds^2 = 
- \frac{r^2}{\ell^2} g(r) dt^2 + \frac{r^2f(r)}{4} [ d\psi + \cos \theta d\phi ]^2 + \frac{\ell^2}{r^2} \frac{dr^2}{f(r)h(r)} + \frac{r^2}{4} d\Omega_2^2 ,
\label{eqn:EH5D_our_metric}
\end{equation}
 where $d\Omega_2^2 = [d\theta^2 + \sin^2\theta d\phi^2]$ is the unit 2-sphere with function solutions
\begin{equation}
\begin{array}{cccc}
f(r) = 1-\frac{\dst r_s^4}{\dst r^4}, & g(r) = 1+\frac{\dst \ell^2}{\dst r^2}, & h(r) = 1+\frac{\dst \ell^2}{\dst r^2}, & \Lambda = -\frac{\dst 6}{\dst \ell^2} .
\end{array}
\label{eqn:EH5D_our_metric_fgh}
\end{equation}

The GB version of the EHS \cite{Wong2012-GB_soliton} has the  large-$r$ (expanded around $r=\infty$)  power-series solution 
\begin{equation}
\begin{array}{lccl}
{} & f_s(r,\alpha) &=& 1 + \frac{\dst a_4}{\dst r^4} - \frac{\dst a_4 (4\alpha a_4 - b_4 \ell^2)}{\dst 2(4\alpha-\ell^2)r^8} - \frac{\dst a_4 \ell^2(12\alpha a_4 - 48\alpha b_4 + 5 b_4\ell^2)}{\dst 15(4\alpha-\ell^2)r^{10}} + {\cal O}\left(\frac{\dst 1}{\dst r^{10}}\right) , \\[6pt]
{} & g_s(r,\alpha) &=& 1 + \frac{\dst \ell^2}{\dst r^2} + \frac{\dst b_4}{\dst r^4} - \frac{\dst b_4 (4\alpha b_4 - \ell^2 a_4)}{\dst 2(4 \alpha - \ell^2)r^8} + \frac{\dst a_4 \ell^2 (252\alpha a_4 - 5 b_4 \ell^2 + 48 \alpha b_4)}{\dst 45(4\alpha - \ell^2)r^{10}} + {\cal O}\left(\frac{\dst 1}{\dst r^{10}}\right) , \\[6pt]
{} & h_s(r,\alpha) &=& 1 + \frac{\dst \ell^2}{\dst r^2} + \frac{\dst b_4}{\dst r^4} - \frac{\dst b_4 (-7\ell^2 a_4 + 12 \alpha b_4)}{\dst 6(4 \alpha - \ell^2)r^8} + \frac{\dst a_4 \ell^2 (756\alpha a_4 - 5 b_4 \ell^2 + 336 \alpha b_4)}{\dst 45(4\alpha - \ell^2)r^{10}} + {\cal O}\left(\frac{\dst 1}{\dst r^{10}}\right) , \\[10pt]
{} & \Lambda &=& -\frac{\dst 6}{\dst \ell^2} + \frac{\dst 12\alpha}{\dst \ell^4} ,
\end{array}
\label{eqn:GB_EH_soliton}
\end{equation}
where we have two main free parameters $a_4 \equiv -r_s^4 + \Delta a_4(\alpha)$ and $b_4 \equiv \Delta b_4(\alpha)$ with $\{ \Delta a_4(\alpha), \Delta b_4(\alpha) \}$ being the contributions induced from GB gravity.
The range of the radial component that the soliton solution admits is $r \in (r_s>0,\infty)$ where $r_s$ is the soliton edge radius; the space inside the `bubble of nothing' edge is non-integrable.
The conditions that must be met at the bubble edge are $f_s(r_s,\alpha)=0$, $g_s(r_s,\alpha) > 0$, and $h_s(r_s,\alpha) > 0$.
Note that the soliton is horizonless because $g_s$ and $h_s$ are non-zero at $r=r_s$.

To ensure continuity with the Einsteinian EHS solution founded by Clarkson and Mann \cite{Clarkson2006-5DSoliton} in the $\alpha \rightarrow 0$ limit,  we look only at the $\Delta a_4(\alpha) = \Delta b_4(\alpha) = 0$ solution -- one class of an infinitely countable set of soliton solutions. In general, both $\Delta a_4(\alpha)$ and $\Delta b_4(\alpha)$ can take non-zero and discrete numerical values 
as shown  by a trial-and-error extrapolation from infinity towards the soliton `bubble' edge $r_s$ so that the boundary conditions are satisfied \cite{Wong2012-GB_soliton}.  We consider this more restrictive $\Delta a_4(\alpha)=\Delta b_4(\alpha)=0$ class of solutions as
we are mainly interested in the small-$\alpha/\ell^2$ (close-to-Einsteinian) physics; it is possible to generalize our results to 
$\Delta a_4(\alpha)\neq 0 \neq \Delta b_4(\alpha)$.

The conical singularities of the Einsteinian EHS can be removed by restricting the angular periodicity of $\psi_s$ to $\eta_s = 4 \pi / k$ for integers $k \ge 3$ so that its metric regularity condition $r_s^2 = \ell^2 \left( k^2/4 - 1 \right)$ can be satisfied.
Its GB analogue has so far been studied using a near-$r_s$ solution  \cite{Wong2012-GB_soliton}. However the analytic connection between the near-$r_s$ parameters and our required arge-$r$ parameters $a_4$ and $b_4$ is not yet clear.  As we are interested
in the {\it qualitative} behaviour of phase transitions,  we shall employ
the Einsteinian metric regularity condition $r_s^2 = \ell^2 \left( k^2/4 - 1 \right)$ with $k=5$, as we expect that in the small-$\alpha$ approximation this will only modify quantitative details.

\subsection{Gauss-Bonnet Orbifolded AdS Schwarzschild Black Hole}
\label{subsec:GB_OSAdS}
The 5-dimensional static and uncharged GB OSAdS (with manifold curvature of $-1$) has the following solution functions
\begin{equation}
\begin{array}{lccl}
{} & f_b(r,\alpha) &=& 1 , \\[6pt]
{} & g_b(r,\alpha) &=& \frac{\dst \ell^2}{\dst r^2} + \frac{\dst 1}{\dst 4\alpha\ell^2}\left[ 1 - \sqrt{1+8\alpha \left( \frac{\dst \mu}{\dst r^4} - \frac{\dst 1}{\dst \ell_{\text{eff}}^2} \right)} \right] \\[6pt]
{} & {} &=& \frac{\dst \ell^2}{\dst \ell_{\text{eff}}^2} + \frac{\dst \ell^2}{\dst r^2} - \frac{\dst \mu \ell^2}{\dst r^4} + \frac{\dst 2\alpha\ell^2}{\dst r^4}\left( \frac{\dst r^2}{\dst \ell_{\text{eff}}^2} - \frac{\dst \mu}{\dst r^2} \right)^2 + {\cal O}\left( \alpha^2 \right) , \\[12pt]
{} & h_b(r,\alpha) &=& g_b(r,\alpha) , \\[10pt]
{} & \Lambda &=& -\frac{\dst 6}{\dst \ell^2} + \frac{\dst 12\alpha}{\dst \ell^4} ,
\end{array}
\label{eqn:GB_OSAdS}
\end{equation}
where $\ell_{\text{eff}}^2 = \ell^2 \left[ 1-2\alpha/\ell^2 \right]^{-1}$ and $\mu$ is the black hole mass.
The range of the radial component that the black hole solution admits is $r \in (r_b>0,\infty)$ where $r_b$ is the black hole (event horizon) radius.
The constraints imposed on the black hole horizon is $g(r_b)=h(r_b)=0$.
By simple algebraic manipulations, this constraint gives the mass parameter $\mu = \mu(r_b,\ell^2,\alpha)$ in terms of the black hole radius as

\begin{equation}
\mu = \frac{r_b^4}{\ell_{\text{eff}}^2} + r_b^2 + 2\alpha .
\label{eqn:mu_GB_blackhole}
\end{equation}

The black hole has a conical singularity at the black hole horizon radius $r_b$ which can be eliminated upon Wick-rotating the time coordinate by $t \rightarrow i \tau$ so that new time coordinate $\tau$ has a period of
\begin{equation}
\beta_b = \frac{2\pi r_b \left( r_b^2 + 4\alpha \right)}{r_b^2 + \frac{2 r_b^4}{\ell^2} \left( 1-\frac{2\alpha}{\ell^2} \right)} .
\label{eqn:beta_b_GB_blackhole}
\end{equation}

This Wick-rotation forces the path integral over all matter fields on the Euclidean section periodic to $\tau$ to be equal to the partition function of the canonical ensemble of matter fields.
From this, we can derive that the black hole temperature $T_b$ is equal to the inverse period $\beta_b^{-1}$.
Although speaking in the language of the black hole temperature is more physical, the black hole radius is actually a more convenient parameter for defining phase transitions because it turns out that a temperature interval $\left( T_b^-, T_b^+ \right)$ does not uniquely determine a phase region, whereas a black hole radius interval $\left( r_b^-,r_b^+ \right)$ does.

\subsection{Orbifolded AdS}
\label{subsec:GB_OSAdS}
The metric functions for 5-dimensional orbifolded thermal AdS are
\begin{equation}
\begin{array}{lccl}
{} & f_a(r,\alpha) &=& 1 , \\[6pt]
{} & g_a(r,\alpha) &=& 1 + \frac{\dst \ell^2}{\dst r^2} , \\[12pt]
{} & h_a(r,\alpha) &=& g_a(r,\alpha) , \\[10pt]
{} & \Lambda &=& -\frac{\dst 6}{\dst \ell^2} + \frac{\dst 12\alpha}{\dst \ell^4} .
\end{array}
\label{eqn:GB_OAdS}
\end{equation}

This is the pure AdS space that corresponds to the field theory `vacuum' and is considered the background metric for orbifolded asymptotically AdS spacetimes to converge towards in the $r \rightarrow \infty$ limit.
The range of $r$ here is the whole space $r \in \left( r_a=0,\infty \right)$ unlike the EHS and OSAdS where spacetime is only defined on $r \in \left( r_{s/b}>0,\infty \right)$.

\section{Finding Phase Transitions and Evolutionary Tracks}
\label{sec:phase_transitions_and_evolutionary_tracks}

To study the  tendencies for phase transitions between two states $i \in \mathcal{S}$ and $j \in \mathcal{S}$, we look to the sign of the relative Euclidean action $I_{ij} \equiv \mathcal{I}_{i} - \mathcal{I}_{j}$ calculated using \eqref{eqn:GB_action_final} up to linear order in $\alpha$.
The relative action $I_{ij}$ will be finite only if the metrics are `lined up' by matching the angular/time coordinates in the $r \rightarrow \infty$ limit.
This is done by requiring the angular periods of $\psi_b$ and $\psi_a$ to be matched up with the soliton periodicity $\eta_s$, and the Wick-rotated timelike coordinate periods $\beta_s$ and $\beta_a$ to be matched up with the Wick-rotated period of the black hole $\beta_b$.
In practise, this matching procedure is performed by evaluating $I_{ij}$ out to some large but finite radius $R$, imposing the periodicity-matching conditions to `line up' the metrics, then finish with an $R \rightarrow \infty$ limit. 
Explicitly, the matching conditions at radius $R$ are
\begin{equation}
\begin{array}{cllccll}
\eta_b &=& \sqrt{f_s(R,\alpha)} \eta_s, & {} & \eta_a &=& \sqrt{f_s(R,\alpha)} \eta_s, \\[10pt]
\beta_s &=& \sqrt{\frac{\dst g_b(R,\alpha)}{\dst g_s(R,\alpha)}} \beta_b, & {} & \beta_a &=& \beta_b. \\[10pt]
\end{array}
\label{eqn:matching_conditions}
\end{equation}

\subsection{Phase regions $i \rightarrow j \rightarrow k$}

We define a {\it phase region} as a region in parameter space (corresponding to a range $\left( r^-_{\partial},r^+_{\partial} \right)$  of black hole radii) for which one of the  actions $\{ \mathcal{I}_{s}, \mathcal{I}_{a}, \mathcal{I}_{b} \}$ of
EHS, OSAdS, and OAdS respectively is smallest.
Hence in each phase region we will have a  specific ordering of $\{ \mathcal{I}_{s}, \mathcal{I}_{a}, \mathcal{I}_{b} \}$ that expresses the respective stability of EHS, OSAdS, and OAdS (and therefore the direction of quantum tunneling between them).   We employ the arrow-notation $i \rightarrow j \rightarrow k$ for $i,j,k \in \mathcal{S}$ to describe the $3! = 6$ different types of  $\mathcal{I}_{i} < \mathcal{I}_{j} < \mathcal{I}_{k}$ phase regions deduced by calculating $I_{ij} \equiv I_i-I_j$ etc. (where a positive value  indicates $i$ tunneling into $j$).  Note that the sign of each $I_{ij}$, $I_{jk}$, and $I_{ik}$ is positive or negative everywhere within a phase region. The boundary of a phase region is where one of  $I_{ij}$, $I_{jk}$, and $I_{ik}$ vanish.

In our arrow-notation, the arrows point in the direction of tunneling to the more-stable state. Hence for $i \rightarrow j \rightarrow k$, the spacetime  $i$ is the least energetically stable state (largest action) and $k$ is the most energetically stable state (smallest action).
By following the arrows, a phase region described by $i \rightarrow j \rightarrow k$ tells an evolutionary story for different initial states:
\begin{enumerate}
\item[] (Pi1) $i$ will tunnel into $j$, and later on tunnel into $k$.
\item[] (Pi2) $i$ will tunnel into $k$ directly (bypassing the intermediate state $j$). 
\item[] \ (Pj) $j$ will tunnel into $k$ (it cannot tunnel into $i$).
\item[] \ (Pk) $k$ will remain the same (it cannot tunnel into neither $j$ nor $k$).
\end{enumerate}
Hence the least stable state $i$ has two tunnelling options, the state $j$ of intermediate stability has one tunnelling option, and the 
stable state $k$ does not change. The relative action $I_{ij}$ evaluated inside a phase region yields the tunnelling amplitude from state $i$ to state $j$ with tunnelling rate $\Gamma_{i \rightarrow j} \sim \exp [ -I_{ij} ]$. Clearly 
 process (Pi2) is less probable than (Pi1) since $\Gamma_{i \rightarrow k} \sim \exp [ -I_{ij} ] \cdot \exp [ -I_{jk} ] < \exp [ -I_{ij} ] \sim \Gamma_{i \rightarrow j}$;  more-excited states prefer to tunnel to less-excited states via intermediate states if possible.
Therefore the most relevant processes will usually be just (Pi1), (Pj), and (Pk) in a phase region with $i \rightarrow j \rightarrow k$.

\subsection{Phase boundaries $r^{ij}_{\partial}$}

\begin{figure}[htb]
\begin{center}
 \includegraphics[width=0.6\textwidth]{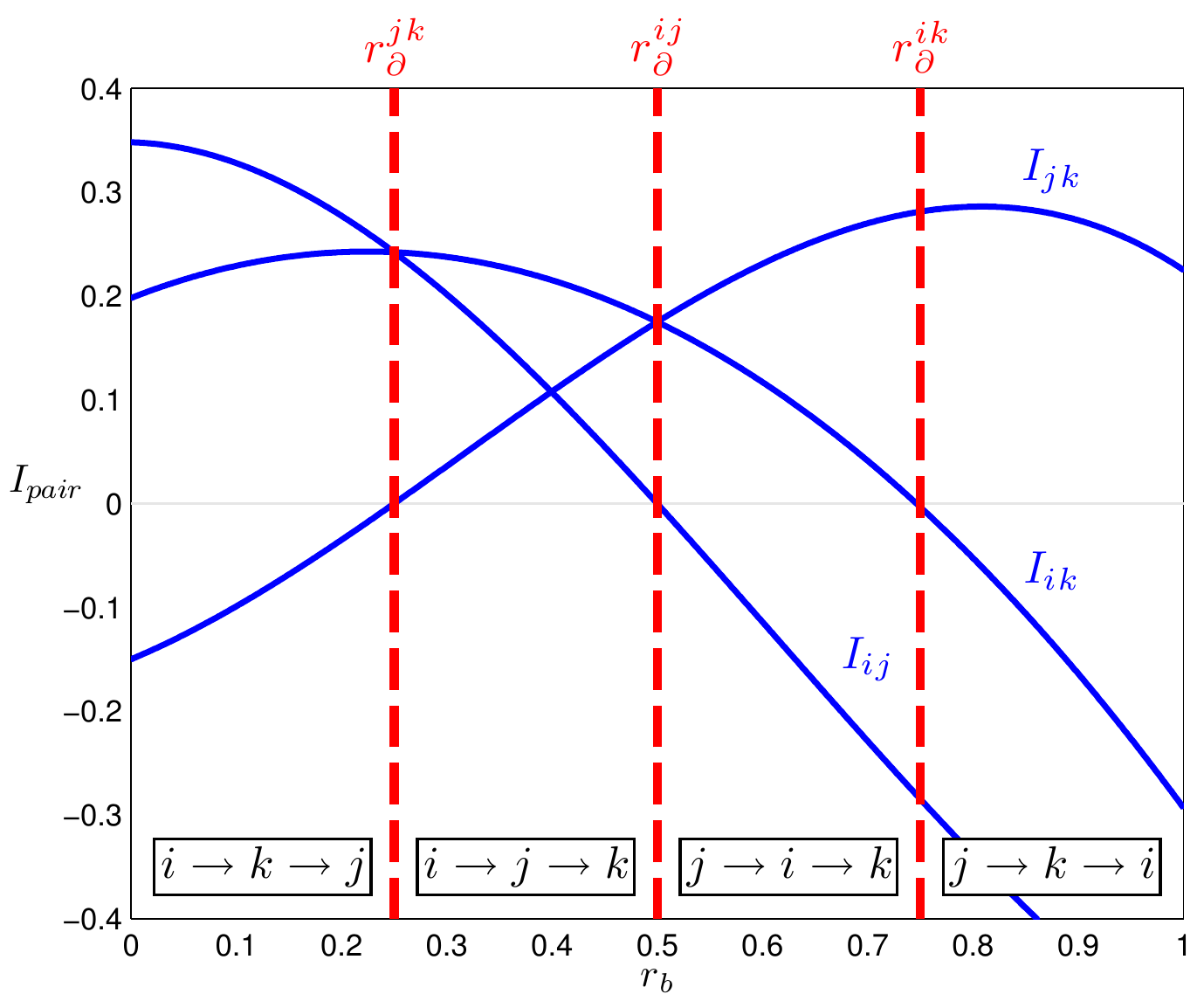}
 \caption{An illustrative plot of the three pairwise relative actions $I_{ij}$, $I_{jk}$, and $I_{ik}$ plotted as a function of $r_b$ (blue solid lines) to form phase region (using $i \rightarrow j \rightarrow k$ notation) and phase boundaries (red dashed lines) for some chosen $\alpha$. Note that a phase boundary is formed whenever there is an $I_{\text{pair}}=0$.}
 \label{fig:pairwise_actions}
\end{center}
\end{figure}

A phase boundary $r_b = r^{ij}_{\partial}$ ($i,j \in \mathcal{S}$ distinct) is where  $I_{ij}(r^{ij}_{\partial})=0$.  A trajectory in parameter
space that crosses such a boundary switches the sign of $I_{ij}$ -- essentially it 
reverses the tunneling direction between {\it exactly one} pair of states $\{ i, j \}$. For example, if we start in a phase region $i \rightarrow j \rightarrow k$ and cross a phase boundary, we either arrive in phase region $j \rightarrow i \rightarrow k$ (after crossing $r^{ij}_{\partial}$) or $i \rightarrow k \rightarrow j$ (after crossing $r^{jk}_{\partial}$).
A quick illustration of crossing phase boundaries  is shown in Fig.~\ref{fig:pairwise_actions}.

One logically remaining possibility is a ``triple point":  a boundary where the actions of EHS, OSAdS, and OAdS are equal.  This would
involve a transition $i \rightarrow j \rightarrow k$ into $k \rightarrow j \rightarrow i$ (requiring $I_{ij}(r^{ik}_{\partial})=0$ and $I_{jk}(r^{ik}_{\partial})=0$ ).  We find no such examples in our investigations. We also note   that a phase region can be bounded by two phase boundaries $r^{-}_{\partial} < r_b < r^{+}_{\partial}$ from the left and the right; but it can also have only one phase boundary when either $r^{-}_{\partial} = 0$ or $r^{+}_{\partial}=\infty$, or even no phase boundary at all if both $r^{-}_{\partial} = 0$ and $r^{+}_{\partial}=\infty$.

\section{Results and Discussion}
\label{sec:resuts_and_discussion}

We discuss here the results of computing the relative actions, the phase diagrams they depict, and their implications.
Given our constraints and assumptions, we can fully describe the phase transitions with \emph{two} parameters: the GB coefficient $\alpha$, and the radius of the black hole $r_b$ (with numerical units of $\ell^2$ and $\ell$ respectively).

By evaluating each $I_{ij} \equiv \mathcal{I}_i - \mathcal{I}_j$ using \eqref{eqn:GB_action_final}, the three $\{as,bs,ba\}$-pairs for the relative action are
\begin{equation}
\begin{array}{ccc}
I_{as} = \frac{\dst \pi \beta_b r_s^4}{\dst 8 \ell^2 k G} + \alpha \frac{\dst \pi \beta_b \left( 48\ell^2 + 56 r_s^2 \right)}{\dst 8 \ell^2 k G} , \\[12pt]
I_{bs} = \frac{\dst \pi \beta_b \left( r_b^4 r_s^2 - 2 r_b^8 + \ell^2 r_b^4 \mu \right)}{\dst 8 \ell^2 r_p^4 k G} + \alpha \frac{\dst \pi \beta_b \left( 24r_b^8 - 8\ell^2 r_b^4 \mu + 48\ell^4 r_b^4 - 12 \ell^4 \mu^2 + 56 \ell^2 r_b^4 r_s^2 \right)}{\dst 8 \ell^4 r_b^4 k G}, \\[12pt]
I_{ba} = I_{bs} - I_{as} 
\end{array}
\label{eqn:I_action_results}
\end{equation}
 to linear order in $\alpha$.

\begin{figure}[h]
\begin{center}
 \includegraphics[width=0.65\textwidth]{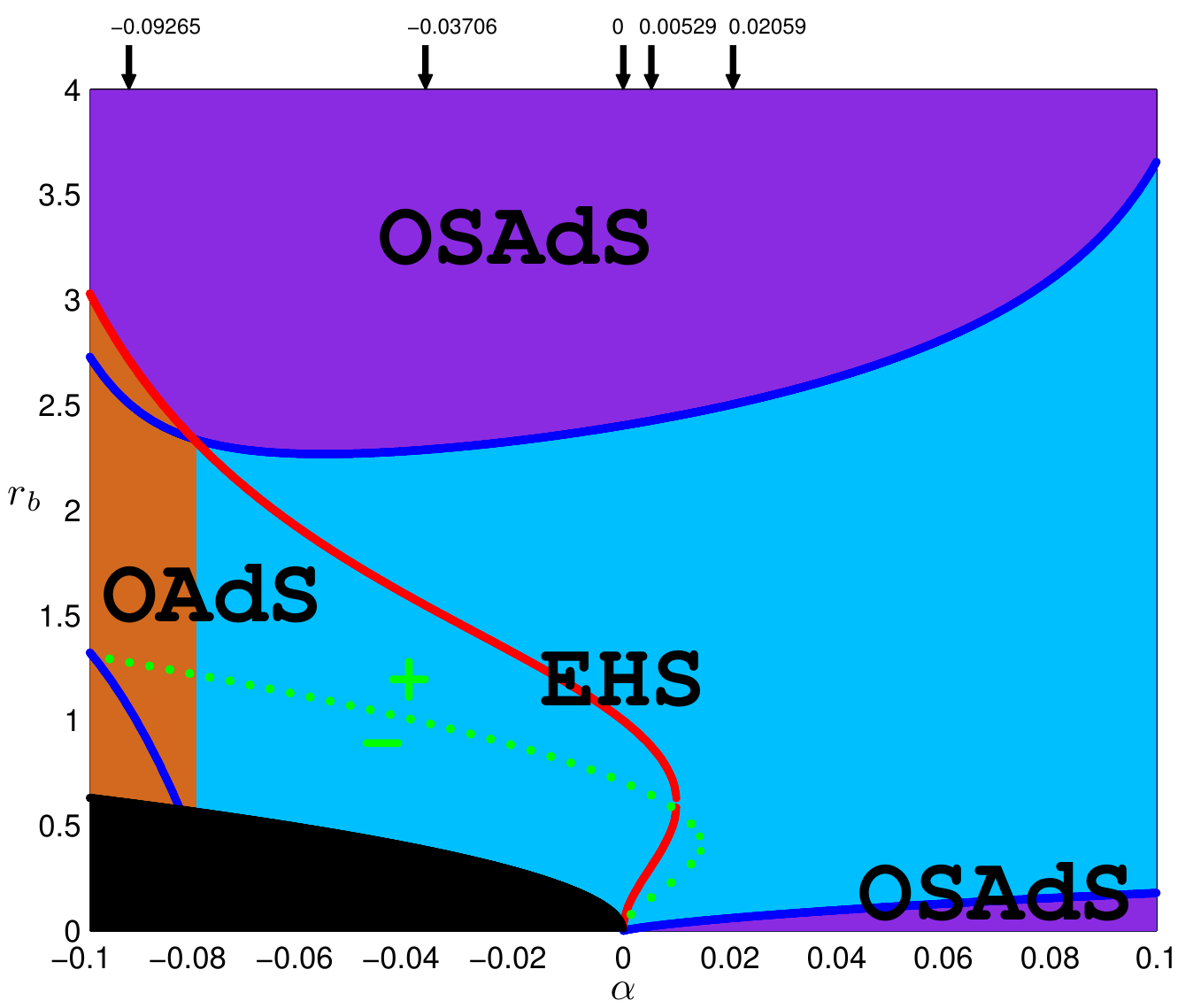}
 \caption{The $\alpha$-$r_b$ phase diagram. The light blue-, purple-, brown- colored regions represent the EHS, OSAdS, OAdS ground states, respectively. The solid blue and red curves are the radii $r_{\partial}$ of the OSAdS-EHS $r^{bs}_{\partial}$ and the OSAdS-OAdS (Hawking-Page) $r^{ba}_{\partial}$ phase boundaries. The green dotted line is {\it not} a phase boundary, instead it indicates where the specific heat capacity of the black hole switches signs (with $+$ and $-$ labels showing the sign of $C_b$). The black region is an inaccesible part of the phase diagram which the OSAdS cannot exist in. The 5 arrows at the top of the phase diagram are the representative values of $\alpha$ noted in the text.}
 \label{fig:r_sab_alpha}
\end{center}
\end{figure}

Given a value of $\alpha$, each phase region is described uniquely by a black hole radius interval $\left( r^-_{\partial}, r^+_{\partial} \right)$ associated with an  $I(r^-_{\partial} < r_b < r^+_{\partial},\alpha)$ of constant sign.
Any choice of   black hole radius $r_b$ will pin-point the phase we are in.
Every phase region has its own ground state $i \in \mathcal{S}$ for which all other states in $\mathcal{S}$ will ultimately evolve into. 
This is illustrated by the coloured ground-state regions and $r^{ij}_{\partial}$ phase boundaries on the $\alpha$-$r_b$ phase diagram in Fig~\ref{fig:r_sab_alpha}.
The green dotted line is {\it not} a phase transition, rather, it indicates where the black hole specific heat capacity $C_b = -\beta_b^2 \partial^2 I_b/\partial_{\beta_b}^2$ diverges and switches signs.
This first-order transition lies exactly where the black hole temperature reaches a local extremum.

We now explore the main physics of this $\alpha$-$r_b$ phase diagram by choosing five $\alpha$ values -- $\alpha = { -0.09265,\ -0.03706,\ 0, \ +0.00529, \ +0.02059 }$ -- that  represent its salient features. 

\subsection{The Einsteinian Phase Diagram: $\alpha = 0$}
\label{subsec:phase_diagram_alpha_zero}
\begin{figure}[htb]
\begin{center}
 \includegraphics[width=0.6\textwidth]{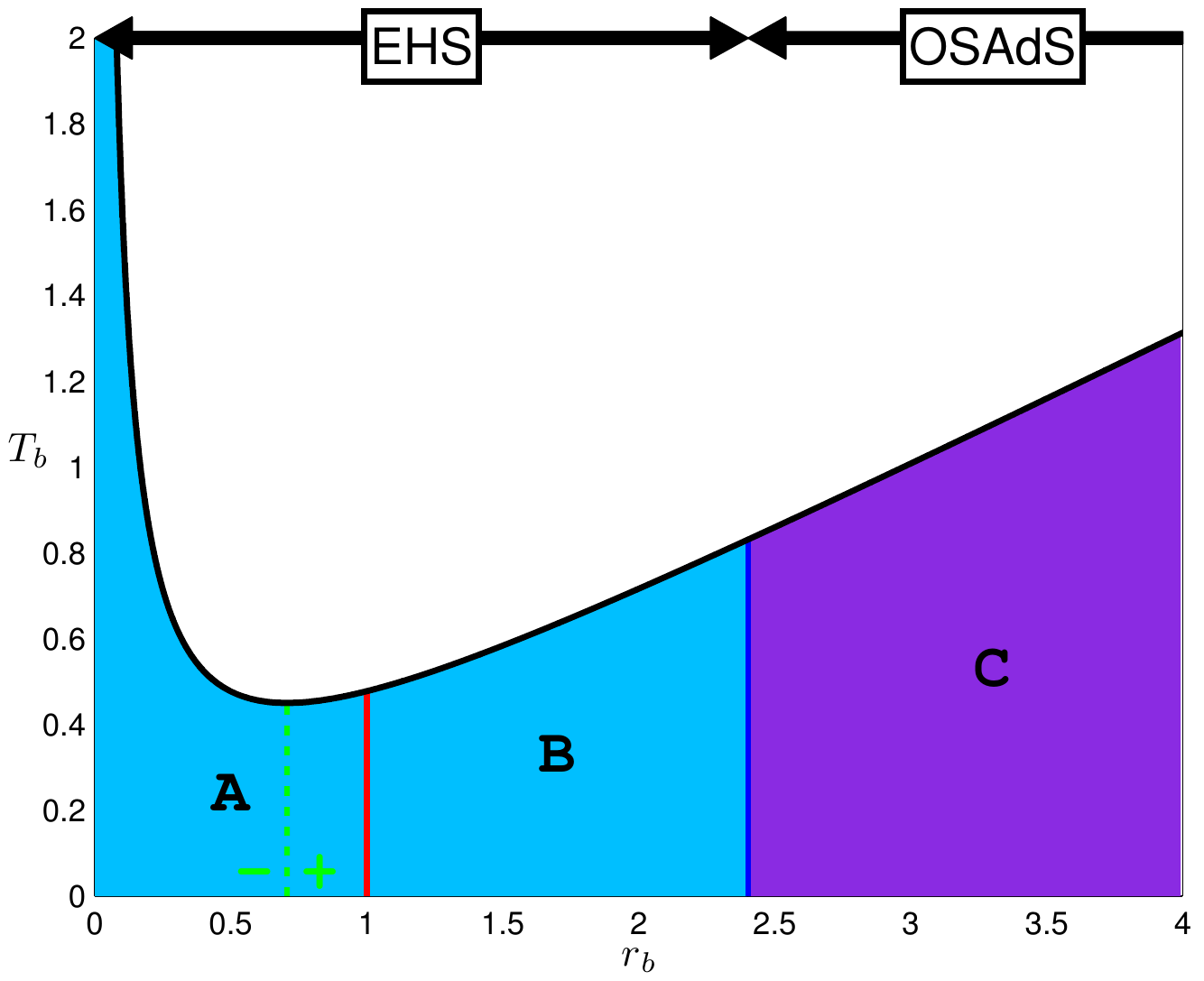}
 \caption{The $\alpha=0$ phase diagram corresponding to traversing up the $\alpha=0$ vertical line on the $\alpha$-$r_b$ phase diagram in Fig~\ref{fig:r_sab_alpha}. The black curve is the temperature of the GB black hole as calculated from Eqn \eqref{eqn:beta_b_GB_blackhole}. The alphabetical labelling of different evolutionary tracks are tabulated in Table~\ref{tab:evolution_table}. The solid blue and red curves are the radii $r_{\partial}$ of the OSAdS-EHS $r^{bs}_{\partial}$ and the OSAdS-OAdS (Hawking-Page) $r^{ba}_{\partial}$ phase boundaries. The green dotted line is {\it not} a phase boundary, instead it indicates where the specific heat capacity of the black hole switches signs (with $+$ and $-$ labels showing the sign of $C_b$). The ground states are labeled at the top.}
 \label{fig:pd_alpha_3}
\end{center}
\end{figure}

As expected, we recover exactly the Einsteinian results of Stotyn and Mann \cite{Stotyn2009-soliton_bh_phasetransitions} when we set $\alpha=0$.
We review here the Einsteinian $\alpha=0$ phase diagram using Fig~\ref{fig:pd_alpha_3} before extending to Gauss-Bonnet gravity.
The alphabetical labelling of the phases regions in Fig~\ref{fig:pd_alpha_3} through Fig~\ref{fig:pd_alpha_5} are catalogued in Table~\ref{tab:evolution_table} using our $i \rightarrow j \rightarrow k$ notation introduced in Section \ref{sec:phase_transitions_and_evolutionary_tracks}. 

The temperature of the Einsteinian OSAdS is infinite in the $r_b \rightarrow 0$ limit because the small black hole is in a near-flat background, and thus follows the usual $T_b \sim r_b^{-1}$ relation.
A   feature of AdS space that differs considerably from flat space is that its large black holes are very hot because $T \sim r_b$  for large $r_b$. 
The minimum and finite temperature that the black hole achieves is at the green dotted line in Fig~\ref{fig:pd_alpha_3} where the black hole specific heat capacity $C_b$ diverges and switches sign.

Semi-classically,  a small black hole for $\alpha=0$ has $C_b < 0$ and so gets hotter with time, emitting Hawking radiation at a faster and faster rate until it `vanishes' to become OAdS which in turn tunnels to the EHS ground state. 
Although having $C_b > 0$ is necessary for black hole stability, it is clearly not sufficient as seen on the right side of region A where it still ultimately ends up as an EHS.
Even past the red $r^{as}_{\partial}$ (Hawking-Page) boundary into region B, the black hole remains unstable by tunnelling into the EHS despite being a little more (semi-classically) stable than it would be in region A.
However, when the black hole gets large enough so that we are in region C by passing through the blue $r^{bs}_{\partial}$ boundary, the OSAdS finally becomes stable enough to be the ground state.
Because the large black hole is stable in $\alpha=0$, we also expect the large black hole to be stable in the later $\alpha \neq 0$ analyses due to the vanishing nature of intrinsic curvature-corrections with increasing black hole size.

\begin{table}[H]
\begin{center}
\begin{tabular}{|c|c|c|c|c|c|}
  \hline
  \multirow{3}{*}{Phase Region} & \multicolumn{2}{c|}{ \multirow{2}{*}{$\alpha<0$} } & \multicolumn{1}{c|}{ \multirow{2}{*}{$\alpha=0$} } & \multicolumn{2}{c|}{ \multirow{2}{*}{$\alpha>0$} } \\
   {} & \multicolumn{2}{c|}{} & \multicolumn{1}{c|}{} & \multicolumn{2}{c|}{}\\ \cline{2-6}
   {} & {\bf -0.09265} & {\bf -0.03706} & {\bf 0} & {\bf +0.00529} & {\bf +0.02059} \\ \hline \hline
  \multicolumn{1}{|c|}{\bf A} & {s $\rightarrow$ b $\rightarrow$ a} & {b $\rightarrow$ a $\rightarrow$ s} & {b $\rightarrow$ a $\rightarrow$ s} & {a $\rightarrow$ s $\rightarrow$ b}& {a $\rightarrow$ s $\rightarrow$ b} \\ \hline
  \multicolumn{1}{|c|}{\bf B} & {b $\rightarrow$ s $\rightarrow$ a} & {a $\rightarrow$ b $\rightarrow$ s} & {a $\rightarrow$ b $\rightarrow$ s} & {a $\rightarrow$ b $\rightarrow$ s}   & {a $\rightarrow$ b $\rightarrow$ s} \\ \hline
  \multicolumn{1}{|c|}{\bf C} & {s $\rightarrow$ b $\rightarrow$ a} & {a $\rightarrow$ s $\rightarrow$ b} & {a $\rightarrow$ s $\rightarrow$ b} & {b $\rightarrow$ a $\rightarrow$ s}   & {a $\rightarrow$ s $\rightarrow$ b} \\ \hline
  \multicolumn{1}{|c|}{\bf D} & {s $\rightarrow$ a $\rightarrow$ b} & {-}                                 & {-}                                 & {a $\rightarrow$ b $\rightarrow$ s}   & {-} \\ \hline
  \multicolumn{1}{|c|}{\bf E} & {-}                                 & {-}                                 & {-}                                 & {a $\rightarrow$ s $\rightarrow$ b} & {-} \\ \hline
\end{tabular}
\caption{A table of $i \rightarrow j \rightarrow k$ evolutionary tracks for reaching the most stable configuration on each phase region where $s = \text{EHS}$, $a = \text{OAdS}$, and $b = \text{OSAdS}$. The alphabetical ordering $A \rightarrow E$ of the regions are in increasing black hole radius $r_b$. The arrows point towards the direction of tunnelling which eventually leads to the ground state.}
\label{tab:evolution_table}
\end{center}
\end{table}

\subsection{The Gauss-Bonnet Phase Diagram: $\alpha < 0$}
\label{subsec:phase_diagram_alpha_negative}

\begin{figure}[htb]
\begin{center}
 \includegraphics[width=0.6\textwidth]{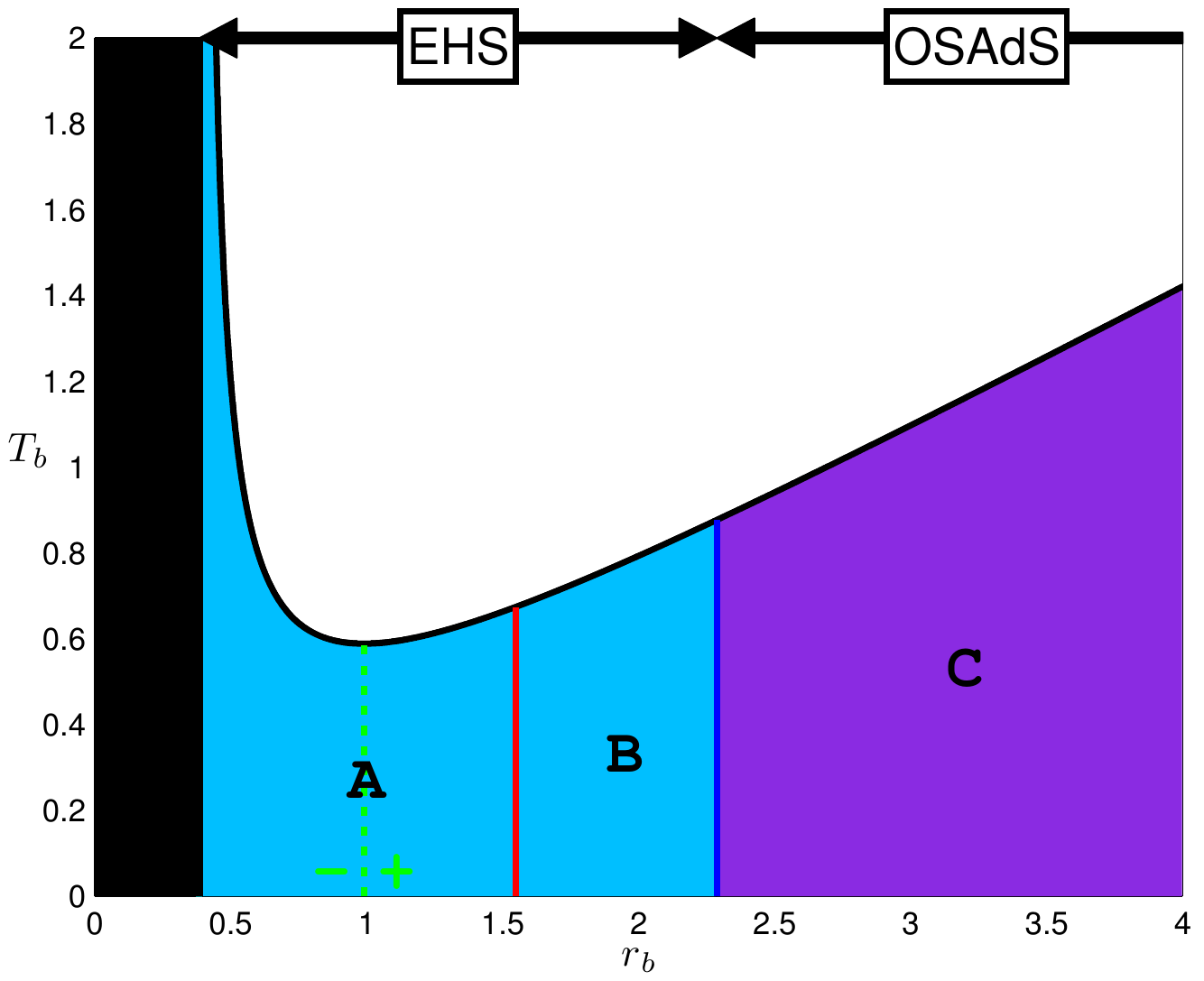}
 \caption{The $\alpha=-0.03706$ phase diagram corresponding to traversing up the $\alpha=-0.03706$ vertical line on the $\alpha$-$r_b$ phase diagram in Fig~\ref{fig:r_sab_alpha}. The black curve is the temperature of the GB black hole as calculated from Eqn \eqref{eqn:beta_b_GB_blackhole}. The alphabetical labelling of different evolutionary tracks are tabulated in Table~\ref{tab:evolution_table}. The solid blue and red curves are the radii $r_{\partial}$ of the OSAdS-EHS $r^{bs}_{\partial}$ and the OSAdS-OAdS (Hawking-Page) $r^{ba}_{\partial}$ phase boundaries. The green dotted line is {\it not} a phase boundary, instead it indicates where the specific heat capacity of the black hole switches signs (with $+$ and $-$ labels showing the sign of $C_b$). The ground states are labeled at the top.}
 \label{fig:pd_alpha_2}
\end{center}
\end{figure}

\begin{figure}[H]
\begin{center}
 \includegraphics[width=0.6\textwidth]{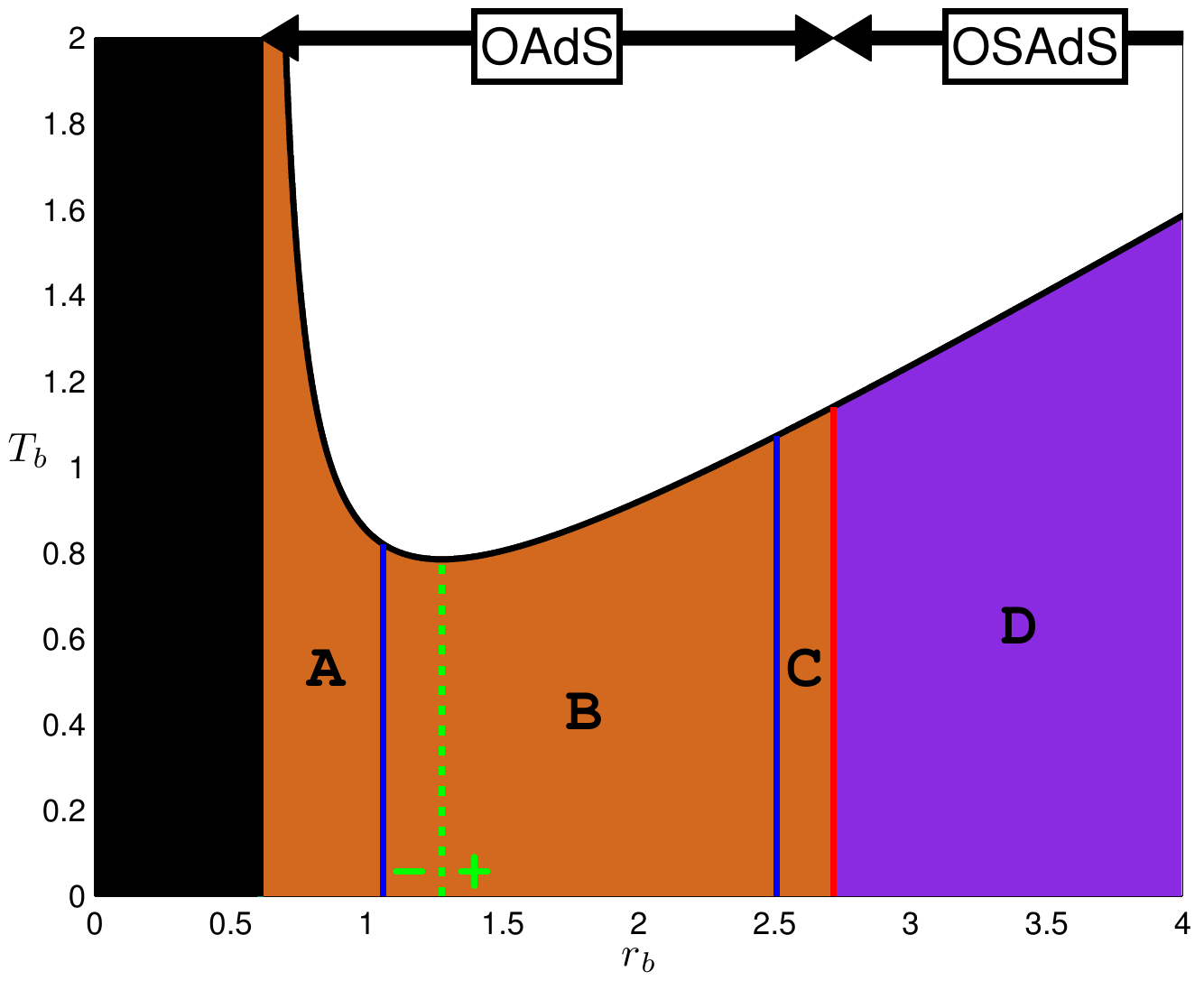}
 \caption{The $\alpha=-0.09265$ phase diagram corresponding to traversing up the $\alpha=-0.09265$ vertical line on the $\alpha$-$r_b$ phase diagram in Fig~\ref{fig:r_sab_alpha}. The black curve is the temperature of the GB black hole as calculated from Eqn \eqref{eqn:beta_b_GB_blackhole}. The alphabetical labelling of different evolutionary tracks are tabulated in Table~\ref{tab:evolution_table}. The solid blue and red curves are the radii $r_{\partial}$ of the OSAdS-EHS $r^{bs}_{\partial}$ and the OSAdS-OAdS (Hawking-Page) $r^{ba}_{\partial}$ phase boundaries. The green dotted line is {\it not} a phase boundary, instead it indicates where the specific heat capacity of the black hole switches signs (with $+$ and $-$ labels showing the sign of $C_b$). The ground states are labeled at the top.}
 \label{fig:pd_alpha_1}
\end{center}
\end{figure}

We now consider the $\alpha<0$ Gauss-Bonnet phase diagram, discussing
 these results in   order of increasingly negative $\alpha$.  We shall start from $\alpha=0^{-}$ using Fig~\ref{fig:pd_alpha_2} then Fig~\ref{fig:pd_alpha_1}.

For any $\alpha < 0$, the OSAdS must have a radii of at least $r_b > r^{\text{min}}_b \equiv 2 \sqrt{-\alpha}$ to be physically relevant as the black hole temperature diverges and turns negative when $r_b$ shrinks past $r^{\text{min}}_b$.
Apart from this lower bound, the small negative $\alpha=-0.03706$ phase diagram in Fig~\ref{fig:pd_alpha_2} is qualitatively the same as the $\alpha=0$ phase diagram: the black hole temperature diverges at both ends, there is a first-order transition in region A, and evoutionary tracks are in the same order by inspection of Table~\ref{tab:evolution_table}.

Although we do not find an $r^{as}_{\partial}$ phase boundary, we can still have OAdS being the ground state for a large negative $\alpha$ by a crossover of the red $r^{ba}_{\partial}$ and blue $r^{bs}_{\partial}$ boundaries at around $\alpha \approx -0.08$.
This {\it implicitly} forms a $r^{as}_{\partial}$ boundary, mediated by the black hole, so that the inner regions A, B, and C have OAdS as the ground state rather than EHS as observed for smaller negative $\alpha$.
Furthermore the evolutionary tracks all change from this $r^{ba}_{\partial}$/$r^{bs}_{\partial}$ crossover, as is clear by inspection of the $\alpha=-0.09265$ and $\alpha=-0.03706$ evolutionary tracks in Table~\ref{tab:evolution_table}.
When we reach the $\alpha \lesssim -0.084$ regime, another blue $r^{bs}_{\partial}$ boundary emerges from $r_b=r^{\text{min}}_b$ and adds an extra phase region.
What we have described here can be seen in Fig~\ref{fig:pd_alpha_1}.

We do indeed achieve a stable large black hole in $\alpha<0$ as predicted, but note that for $\alpha \lesssim -0.084$ we have an $\text{s} \rightarrow \text{a} \rightarrow \text{b}$ evolutionary track at the large $r_b$ limit rather than the $\text{a} \rightarrow \text{s} \rightarrow \text{b}$ track expected from the $\alpha=0$ results.
This could be an indication that our approximations are starting to break down at this so-called large negative $\alpha$.
But if it is true that OAdS is the ground state at this level, this would imply that a large negative $\alpha$ has the ability to lift the energies of both the OSAdS and EHS states above AdS for small black holes.

\subsection{The Gauss-Bonnet Phase Diagram: $\alpha > 0$}
\label{subsec:phase_diagram_alpha_positive}

\begin{figure}[htb]
\begin{center}
 \includegraphics[width=0.6\textwidth]{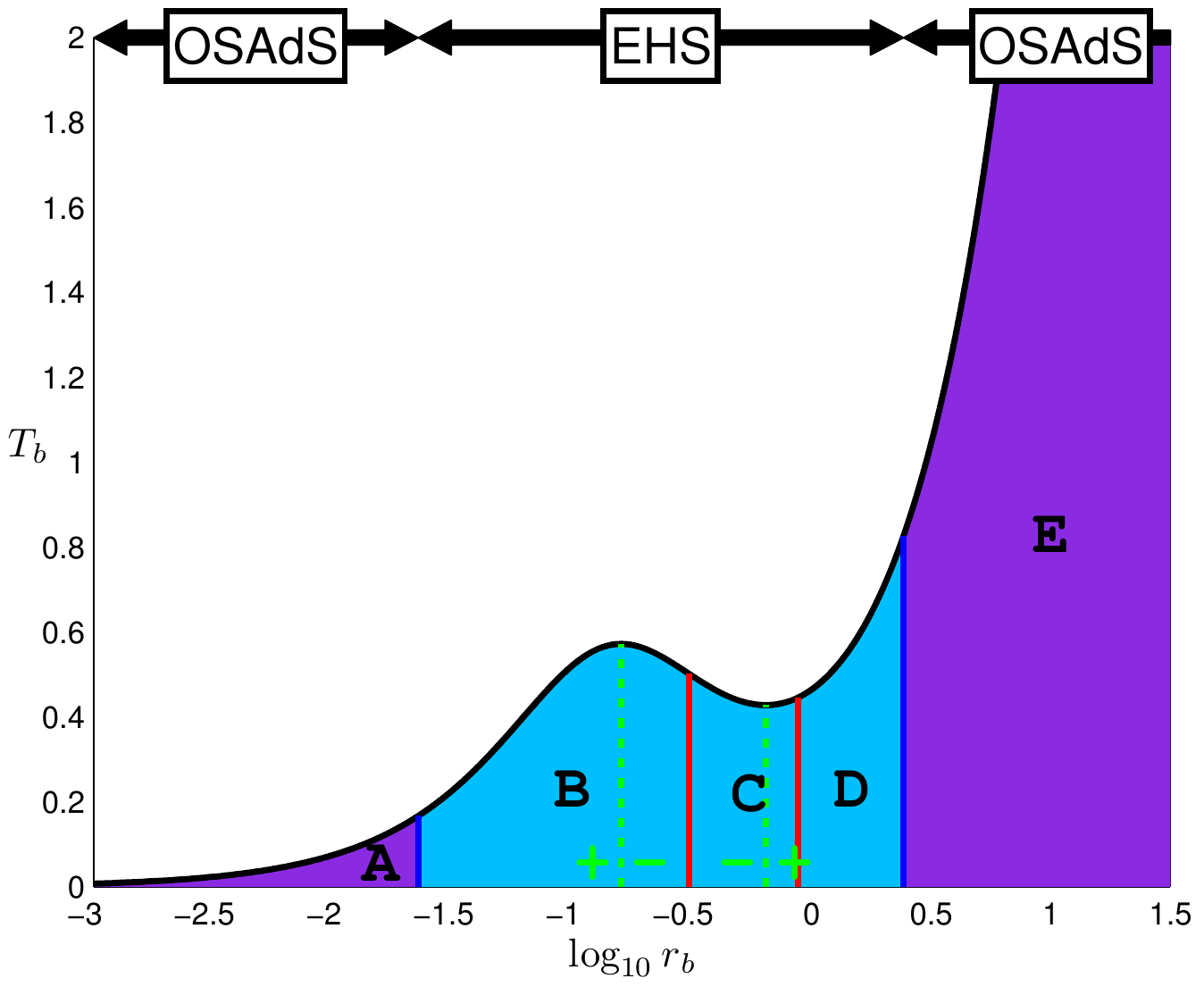}
 \caption{The $\alpha=+0.00529$ phase diagram corresponding to traversing up the $\alpha=+0.00529$ vertical line on the $\alpha$-$r_b$ phase diagram in Fig~\ref{fig:r_sab_alpha}. The black curve is the temperature of the GB black hole as calculated from Eqn \eqref{eqn:beta_b_GB_blackhole}. The alphabetical labelling of different evolutionary tracks are tabulated in Table~\ref{tab:evolution_table}. The solid blue and red curves are the radii $r_{\partial}$ of the OSAdS-EHS $r^{bs}_{\partial}$ and the OSAdS-OAdS (Hawking-Page) $r^{ba}_{\partial}$ phase boundaries. The green dotted line is {\it not} a phase boundary, instead it indicates where the specific heat capacity of the black hole switches signs (with $+$ and $-$ labels showing the sign of $C_b$). The ground states are labeled at the top.}
 \label{fig:pd_alpha_4}
\end{center}
\end{figure}

\begin{figure}[H]
\begin{center}
 \includegraphics[width=0.6\textwidth]{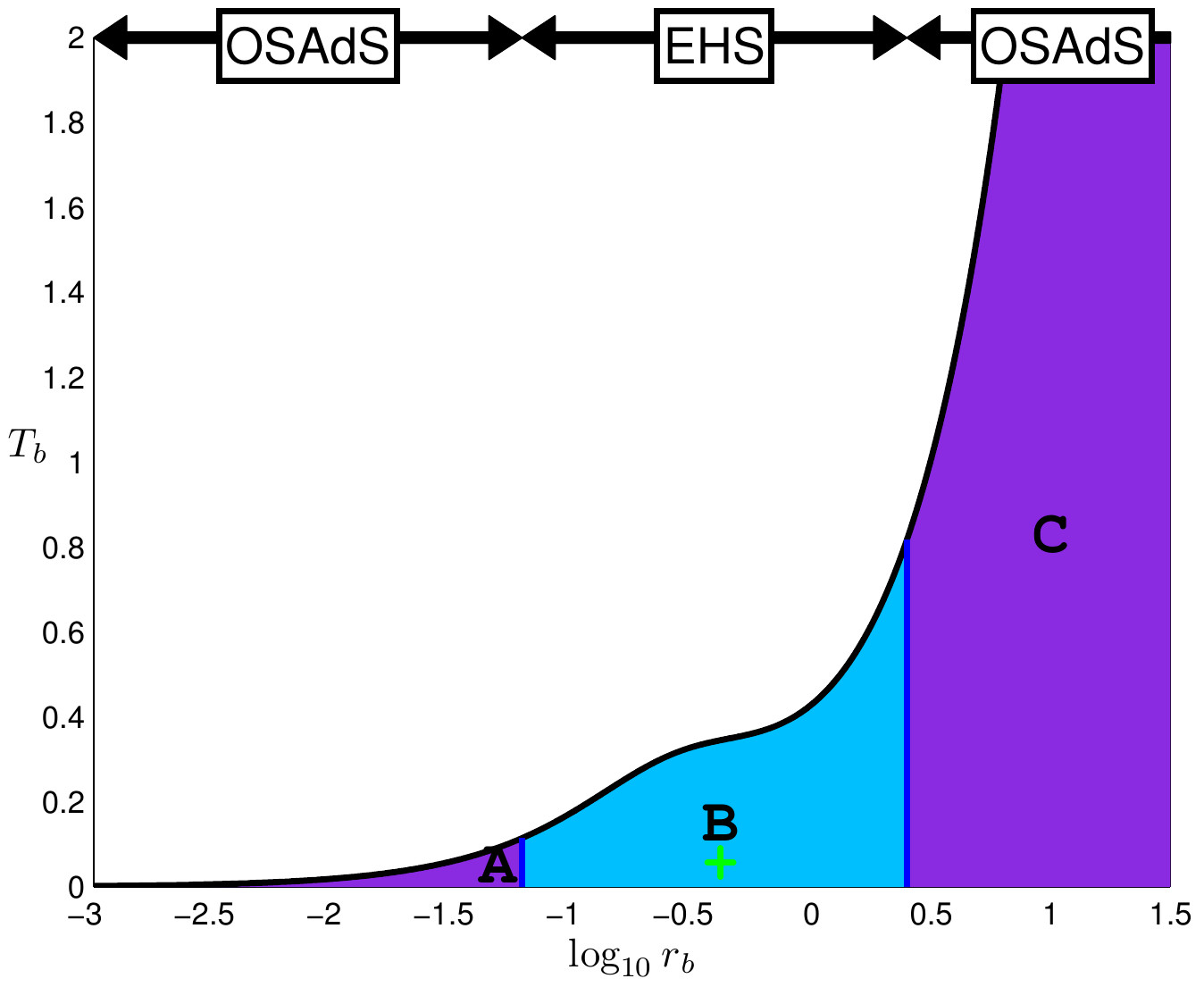}
 \caption{The $\alpha=+0.02059$ phase diagram corresponding to traversing up the $\alpha=+0.02059$ vertical line on the $\alpha$-$r_b$ phase diagram in Fig~\ref{fig:r_sab_alpha}. The black curve is the temperature of the GB black hole as calculated from Eqn \eqref{eqn:beta_b_GB_blackhole}. The alphabetical labelling of different evolutionary tracks are tabulated in Table~\ref{tab:evolution_table}. The solid blue and red curves are the radii $r_{\partial}$ of the OSAdS-EHS $r^{bs}_{\partial}$ and the OSAdS-OAdS (Hawking-Page) $r^{ba}_{\partial}$ phase boundaries.  There is no green dotted line here to locate local extrema in the specific heat capacity because it is positively monotonic throughout all black hole radii (the $+$ label shows the sign of $C_b$).
 The ground states are labeled at the top. }
 \label{fig:pd_alpha_5}
\end{center}
\end{figure}

We now look at the $\alpha>0$ Gauss-Bonnet phase diagram.
We will discuss these results in the order of increasingly positive $\alpha$ starting from $\alpha=0^{+}$ using Fig~\ref{fig:pd_alpha_4} then Fig~\ref{fig:pd_alpha_5}.

One of the immediate features of the $\alpha>0$ phase diagram is how much the topology has changed when juxtaposed with the $\alpha=0$ phase diagram.
For a slightly positive $\alpha = +0.00529$ in Fig~\ref{fig:pd_alpha_4}, we immediately have an extra first-order transition, a red $r^{ba}_{\partial}$ phase boundary, and a blue $r^{bs}_{\partial}$ phase boundary all sprouting from $r_b=0$.
The duplication of these features is a manisfestation of the non-linearity of the problem when accomodating for higher-order curvature correction terms because distinct roots can develop from solving the $r^{ij}_{\partial}$ equation.
Even with the increasing numbers of phase boundaries, we still do not observe an $r^{as}_{\partial}$ boundary and conclude that an $r^{as}_{\partial}$ boundary does not exist in our analysis.

Unlike the $\alpha \le 0$ cases, a zero black hole temperature is attained when the black hole radius shrinks to zero radius for $\alpha>0$.
This comes from the emergence of a new first-order transition boundary which allows the small black hole to have a positive specific heat capacity as it gets smaller.
By doing so, the small black hole is able prevent itself from being `wiped out' by thermal fluctuations by relying on the counteracting effects of its intrinsically-strong curvature-induced quantum fluctuations at low temperatures to keep it stable.
This intuition is verified by the existence of a stable OSAdS in the inner phase region A in both Fig~\ref{fig:pd_alpha_4} and Fig~\ref{fig:pd_alpha_5}.
Also, as expected, the $\alpha > 0$ large black hole is stable: a result consistent with the $\alpha=0$ black hole.
With all this being mentioned, it becomes peculiar that the most unstable black hole configuration for $\alpha>0$ is when it is medium-sized {\it i.e.} an $r_b$ sitting in regions B, C, and, D in Fig ~\ref{fig:pd_alpha_4} or region B in Fig~\ref{fig:pd_alpha_5}.
Note that Cai \cite{Cai2002-GB_BH_AdS} finds the small GB black hole to be stable in $n=5$ as well, but with the added note that this type of black hole stability is supressed in higher $n>5$ dimensions.
This would suggest that the topology of the $\alpha>0$ phase diagram is significantly different from what we have here if this work were to be extended to higher odd $n \ge 7$ dimensions.

Although new exotic behaviour takes place for slightly postive $\alpha$, we find a loss in phase structure for a larger positive $\alpha$.
Explicitly, as we increase from $\alpha=0^{+}$ to $\alpha = 0.02$, the two red $r^{ba}_{\partial}$ (Hawking-Page) phase transitions collide and disappear, consequently eliminating the $\text{b} \rightarrow \text{a} \rightarrow \text{s}$ evolutionary track.
Afterwards, the two first-order transitions collide to become an inflexion point that disappears as well;  all local extrema in $T_b$ are discarded, yielding  a totally positive and smooth specific heat capacity $C_b$ for all black hole radii.
Fig~\ref{fig:pd_alpha_5} shows explicitly the remnants of the phase-removing process associated with having a large positive $\alpha = +0.02059$.
Note that this large positive value of $\alpha$ makes $T_b$ one-to-one with $r_b$, and it is until now that a particular temperature $T_b$ will uniquely identify the phase region we are in because of monotonicity.

\section{Summary and Prospective Research}
\label{sec:summary_and_prospective}

The sign and size of the GB parameter $\alpha$ exploits different physics in GB gravity.
By using a small $\alpha/\ell^2 \ll 1$ approximation, we have constructed an $\alpha$-$r_b$ phase diagram and found extra Hawking-Page and OSAdS-EHS phase transitions emerging for large-enough negative and small-enough positive $\alpha$.
This additional  phase structure is not always the case: a sufficiently small  negative $\alpha$ gives rise to almost the same phase structure as the Einsteinian case, and  phase structure is lost  for sufficiently large positive $\alpha$.

When $\alpha<0$, it is possible for quantum fluctuations to lift the energies of both EHS and OSAdS so that thermal OAdS becomes the ground state.
What makes this more interesting is there is no {\it explicit} $r^{as}_{\partial}$ phase transition to carry this procedure; instead it is {\it implicitly} done by the crossover of the $r^{ba}_{\partial}$ and $r^{bs}_{\partial}$ boundaries at $\alpha \approx -0.08$.
When $\alpha>0$, quantum fluctuations keep small black holes stable by counteracting against the weaker opposing thermal fluctuations actuated by Hawking radiation. 
Given that the small and large black holes are stable, the medium-sized black hole turns out to be the most unstable black hole configuration in $\alpha>0$ as it ends up as the EHS.
Also, the double phase boundaries found from $\alpha=0^+$ can merge together and vanish with increasingly positive $\alpha$.
But despite all the differences between oppositely signed $\alpha$, we find that our finite $\alpha$ large black hole is stable because of its intrinsically small curvature which in turn drives the black hole to behave more like it would in Einsteinian gravity. 

Although the field theory dual to the gravitational $\alpha=0$ phase transitions have been identified as closed-string tachyon condensation, we still await for a more generalized field theory description of our $n=5$ gravitational-side results as means of prediction and verification.
An obvious extension of this work would be to consider a more generalized form of the EHS in GB gravity by exploring the range of values that $a_4$ and $b_4$ can take on.
Exploring this work in dimensions $n \ge 7$ would be intriguing, albeit tedious,  as it would incorporate curvature corrections of at least cubic order which is expected to give different physics as we already know that the higher-dimensional small black hole ceases to be stable for $\alpha>0$.

\acknowledgements
This work was supported by the Natural Sciences and Engineering Research Council of Canada.
We thank Sean Stotyn for useful discussions.

\bibliographystyle{unsrt}
\bibliography{PhaseTransition_references}

\end{document}